\documentclass[a4paper,8pt]{article}
\usepackage[utf8]{inputenc}
\usepackage[T1]{fontenc}
\usepackage{graphicx}
\usepackage{grffile}
\usepackage{longtable}
\usepackage{wrapfig}
\usepackage{rotating}
\usepackage[normalem]{ulem}
\usepackage{amsmath}
\usepackage{textcomp}
\usepackage{amssymb}
\usepackage{capt-of}
\usepackage{hyperref}
\usepackage[sc]{mathpazo} 
\usepackage{microtype} 
\usepackage[english]{babel} 
\usepackage[margin=1.2in]{geometry} 
\usepackage{parskip}
\usepackage[hang, small,labelfont=bf,up,textfont=it,up]{caption} 
\usepackage{booktabs} 
\usepackage{lettrine} 
\usepackage{enumitem} 
\setlist[itemize]{noitemsep} 
\usepackage{abstract} 
\usepackage{titlesec} 
\renewcommand\thesection{\roman{section}} 
\renewcommand\thesubsection{\alph{subsection}.} 
\titleformat{\section}[block]{\large\scshape\centering}{\thesection.}{1em}{} 
\titleformat{\subsection}[block]{\bf\normalsize}{\thesubsection}{0.5em}{} 
\usepackage{fancyhdr} 
\usepackage{titling} 
\usepackage{hyperref} 
\usepackage{subcaption}
\usepackage{float}
\usepackage{tikz}
\usetikzlibrary{shapes,arrows}
\usepackage{amsmath}
\usepackage{csquotes}
\usepackage[ruled,vlined]{algorithm2e} \IncMargin{1.2em}

\author{\\ \textsc{Khan Mohammad Rashedun-Naby}\\ 
\normalsize kmrashedun@gmail.com} 
\date{}
\title{A Peer-to-Peer Distributed Secured Sustainable Large Scale Identity Document Verification System With BitTorrent Network And Hash Function}
\hypersetup{
 pdfauthor={Khan Mohammad Rashedun-Naby},
 pdftitle={A Peer-to-Peer Distributed Secured Sustainable Large Scale Identity Document Verification System With BitTorrent Network And Hash Function},
 pdfkeywords={BitTorrent, Peer-to-Peer, Distributed System, Distributed Database, Hash Function, Identity Document Verification},
 pdfsubject={},
 pdfcreator={Emacs 26.3 (Org mode 9.3.6)}, 
 pdflang={English}}
\begin{document}

\maketitle
\begin{abstract}
\vspace*{1em}
\noindent Verifying identity documents from a large Central Identity Database(CIDB) is always challenging and it get more challenging when we need to verify a large number of documents at the same time. Usually most of the time we setup a gateway server connected to the CIDB and it serve all the identity document verification requests. Though it work well, there are still chances that this model will collapse in high traffic. We obviously can tune the system to be sustainable, but the process is economically expensive. In this paper we propose an economically cheaper way to verify ID documents with private BitTorrent network and hash function.

\vspace*{1em}

\noindent\input{src/00-keywords.org}
\end{abstract}

\section{Introduction}
\label{sec:org46ef95a}
When verifying an individual identity(ID) document from a large \emph{Central Identity Database} or \emph{CIDB} we mostly prefer traditional gateway server based client-server model. Though it work well, the system may face some issues if the load get highly increased. In that case it may need certain upgrades to scale it's operations. But these upgrades are economically quite expensive and resource consuming too. Besides, these public gateway servers has some security concerns too. Like, these open API gateway servers are often a subject to massive DDoS\textsuperscript{\cite{cisa:ddos}} attacks. Obviously by taking some counter measures we can defeat these, but still these measures are expensive.

\emph{BitTorrent\textsuperscript{\cite{cohen2003incentives}}}, a distributed file sharing system which can efficiently share large files over a peer-to-peer network. \emph{Facebook\textsuperscript{\cite{fb_bt}}} and \emph{Twitter\textsuperscript{\cite{tw_bt}}} has made their deployment fast and efficient with it. Even \emph{UK Government\textsuperscript{\cite{uk_govt_bt}}} is using it to distribute their datasets. Our idea is to use this \emph{BitTorrent\textsuperscript{\cite{cohen2003incentives}}} technology to build a private peer-to-peer grid of nodes where every node will have an encrypted version of CIDB in it, which will be referred as \emph{Distributable Identity Database} or \emph{DIDB}. Thus, each of these nodes with DIDB will be able to act as an identity document verification server on demand. As the DIDB will be encrypted, extracting data from DIDB would be impossible. Moreover, we will make it fairly small by encryption and therefore we will be able to store them in average machines. Then these machines will enter the private network as nodes. This will largely cut the expense of setting up and running this identity verification system with no compromise at the other end. Machines outside of the private network, will not be allowed to have the DIDB. But they will be able to send verification requests to the private nodes directly by IP addresses. This distributed identity verification system will also ensure the isolation of the CIDB form the outside world. Hereby, security strength of CIDB will be enhanced. Furthermore, this system will barely need human interactions like security or maintenance, means high autonomy and less chance of human error.
\section{Hash Function}
\label{sec:org02a882c}
Hash function\textsuperscript{\cite{kessler1998overview}} is a special type of cryptography mechanism which generate the output from the input by algorithm only, no key is involved in the process. The length of output is always fixed and the hashing process can not be reversed. Thus, these functions often called \emph{One Way Hash Function\textsuperscript{\cite{Levin_2003}}.} Usually in a hash function, hundreds of one way hashing operations get done sequentially and the results from earlier operations get used in generating the later ones. So, to reverse the hashing process one need to guess hundred percent correct pieces for all the way back. A single wrong piece will give a totally different output and finding these with trial and error has a very large number of combinations to guess. Therefore, we can say these hash functions are completely irreversible. In this paper, we will evaluate our concept with \emph{SHA-256(truncate mode)\textsuperscript{\cite{kelsey2005sha}}}. We choose \emph{SHA-256\textsuperscript{\cite{penard2008secure}}} cause it is battle tested for hash collision with some very good use cases in history, like \emph{Bitcoin\textsuperscript{\cite{nakamoto2008bitcoin}}}, the oldest blockchain network which is doing excellent even though it has reached \emph{4 quadrillion hashes per second\textsuperscript{\cite{bc_hash_rate}}} by 2014.
\section{System Design}
\label{sec:org36abfda}
In our concept we will take a portion of CIDB data and put it in DIDB after encryption. Then the DIDB will be distributed over our private peer-to-peer network. ID document verification requests will be served on demand basis from these peers. And the nodes who will host the DIDB, can do the verification inside them by querying the DIDB clone they have. Now we will explore the technical facts to achieve that.
\subsection{Data Type Selection}
\label{sec:org56d2e87}
The CIDB store various types of data. For example, there are \emph{46} fields in a new voter registration form\textsuperscript{\cite{ec_new_voter_form_2}} of \emph{Bangladesh Election Commission}. But for ID document verification we will use only some common information found in a standard ID document. Like, in a smart NID card\textsuperscript{\cite{dhaka_tribune_smartcard}}, provided by \emph{Bangladesh Election Commission,} we see these information, \emph{(1)ID Serial, (2)Name, (3)Date of Birth, (4)Blood Group, (5)Place of Birth} and \emph{(6)Issue Date}. We will use these to evaluate our concept.

\begin{figure}[htbp]
\centering
\includegraphics[width=1.0\linewidth]{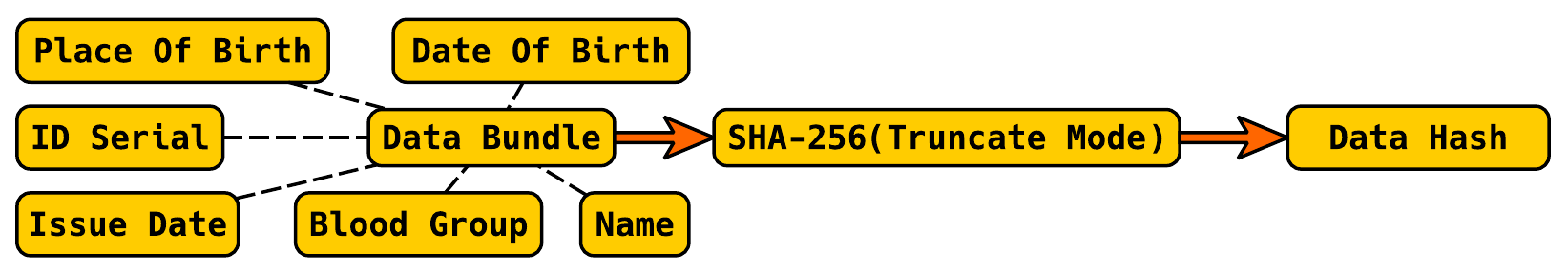}
\caption{Process of creating data hash with multiple types of data}
\end{figure}
We can get these data from the chip or the barcode in the smart card or we can input them manually. Anyway, we will first bundle these data together and then by \emph{SHA-256(truncate mode)\textsuperscript{\cite{kelsey2005sha}}} we will turn the bundled data into hash. This will secure the data and save a lot of space as well. But for indexing them in order to enhance query speed, the hash will be prefixed with birth year-month before going to DIDB. 

\subsection{Disk Space Management}
\label{sec:org47c818f}
We are storing \emph{6 characters} of birth year-month with \emph{40 characters} of data hash. So each block of data will be,
\begin{center}
( 40+6 ) bytes = 46 bytes
\end{center}
Therefore, if the number of identities in a CIDB is \emph{n} then total disk space required for DIDB is,
\begin{center}
( \emph{n} \texttimes{} 46 ) bytes
\end{center}
For example, the population of \emph{Bangladesh} is now a little less than \emph{170 million\textsuperscript{\cite{bd_data}}.} Then for \emph{Bangladesh} we need,
\begin{center}
\Large\(\frac{170000000 \times 46}{1024 \times 1024 \times 1024}\) \normalsize = 7.28294253349 \(\approx\) 7.30 gigabytes
\end{center}
On average \emph{2.7 million\textsuperscript{\cite{bdnews24_new_voter_2018,dhaka_tribune_new_voter_2018,arab_news_new_voter_2018}}} voters get enlisted each year in \emph{Bangladesh}. Here we have added an extra \emph{0.3 million} to this value to minimize any hypothetical error. Then for \emph{3 million} increase in the CIDB each year, our DIDB will grow,
\begin{center}
\Large\(\frac{3000000 \times 46}{1024 \times 1024}\) \normalsize = 131.607055664 \(\approx\) 132 megabytes/year
\end{center}
So after \emph{10 years} the DIDB will grow,
\begin{center}
\Large\(\frac{132 \times 10}{1024}\) \normalsize = 1.2890625 \(\approx\) 1.29 gigabytes
\end{center}
As machines with less than \texttt{500GB} storage capacity is almost a myth now, this \texttt{1.29GB} increase in the DIDB data volume is the very last thing to worry about.
\subsection{DIDB Architecture}
\label{sec:org8229349}
As our application is more performance driven and we have a very low variety of data to store, thus we will put all the ID data one by one like a long string.

\begin{center}
\vspace*{-2em}
\[
\underbrace{\overbrace{198012}^\text{\it{yyyymm}}\overbrace{aaf4c61ddcc5e8a2dabede0f3b482cd9aea9434d}^\text{\it{Identity Data Hash}}}_\text{\it{Single Idenity Record}}
\]
\vspace*{-1em}
\end{center}
And the DIDB engine will know that every \emph{46 characters} represent a single ID data, of which the first \emph{6 characters} are the birth year-month and the rest \emph{40 characters} are the data hash. But storing all the records in a single file has some performance drawbacks. Therefore, it is better to split the DIDB into small chunks and splitting \texttt{7.30GB} in \texttt{4.9MB} of chunk will give-

\begin{center}
\Large\(\frac{7.30\times1024\ MB}{4.9\ MB}\) \normalsize = 1525.55102041 \(\approx\) 1526 files
\end{center}
Here, we will use \emph{LevelDB\textsuperscript{\cite{ghemawat2014leveldb}}} to index those files by birth year-month which will enhance the query speed. To get an idea of the process we can explore \emph{Bitcoin} core file system\textsuperscript{\cite{bc_file_structure}} a bit.
\subsection{Data Validation}
\label{sec:orgf7d5ecb}
We will validate DIDB files by SHA-256\textsuperscript{\cite{sha256sum}} checksum. For \emph{Bangladesh} there would \emph{1526} chunks of files and each file will be \texttt{64 bytes} then,
\begin{center}
\Large\(\frac{1526\times64}{1024\times1024}\) \normalsize = 0.09313964843 \(\approx\) 0.094 megabytes
\end{center}
\subsection{Network Architecture}
\label{sec:orgb9c4453}
Outside machines will be able to send verification requests to private nodes directly by IP addresses. Either by a server or \emph{DHT\textsuperscript{\cite{Galuba2009}}} the IP addresses can be relayed outside.

\begin{figure}[htbp]
\centering
\includegraphics[width=1.0\linewidth]{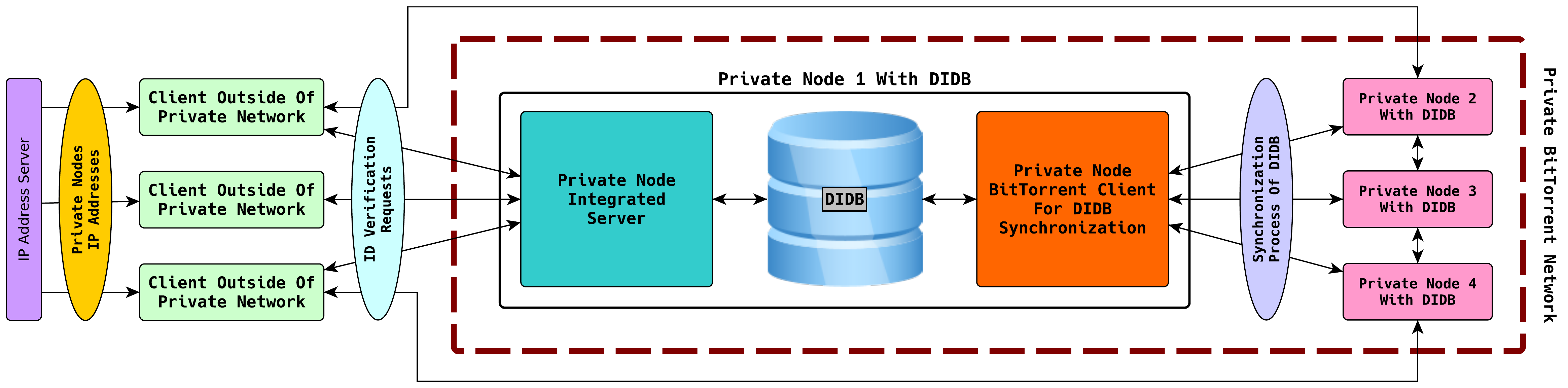}
\caption{A diagram of distributed ID document verification network}
\end{figure}

In a private node client there would be three modules, \emph{(1)DIDB, (2)Private BitTorrent\textsuperscript{\cite{cohen2003incentives}} Client} and \emph{(3)Integrated Server}. The \emph{BitTorrent\textsuperscript{\cite{cohen2003incentives}}} module will synchronize the DIDB and the integrated server will serve the verification requests.
\subsection{Managing DIDB Updates}
\label{sec:org92a8c8c}
We can auto generate DIDB after a specific time period and if the new DIDB is \texttt{10GB} then to get it with \texttt{1MB/s} we may need,
\begin{center}
\Large\(\frac{10\ GB}{1\ MB/s}\) = \(\frac{10240\ MB}{1\ MB/s}\) = \(\frac{10240}{60 \times 60}\) = \normalsize 2.844 \(\approx\) 3.0 hours
\end{center}
So all the private nodes will get it within \texttt{3 hours} on average. For individual updates like a correction of an ID data in CIDB, the DIDB file updated in the main node will be identified by checksum and hereby will be updated.

\subsection{Verification Requests From Outside}
\label{sec:org4732480}
First, a machine outside of the private network, will generate the parameter by hashing all the inputs and prefixing it with the birth year-month. Then it will ask the trusted server for the list of private nodes IP addresses if it does not have it already.

\begin{figure}[htbp]
\centering
\includegraphics[width=1.0\linewidth]{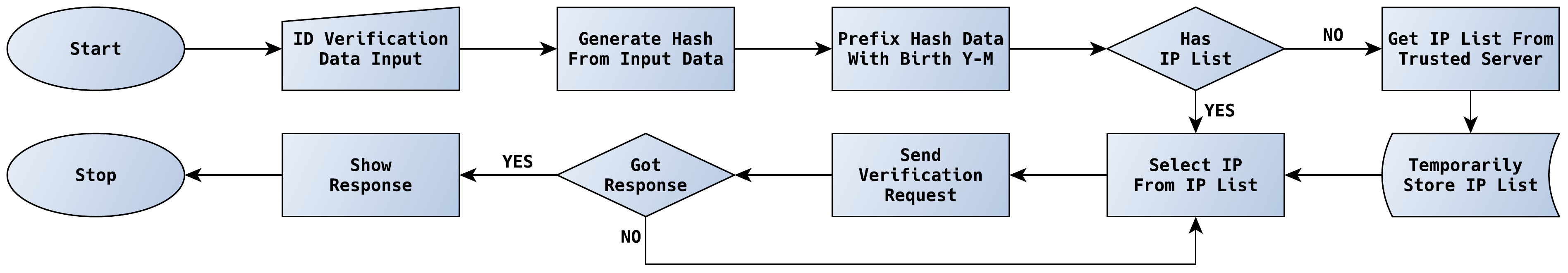}
\caption{A verification request life-cycle}
\end{figure}

After that it will send the verification request with the parameter to a private node IP address selected from the list. On failure of the private node we may try any other IP from the list.

\subsection{Concept Implementation}
\label{sec:orgc96c8a9}
Our concept is quite easy to implement. We just need the client application and the DIDB generation module to generate the DIDB form CIDB. No further network or hardware modifications are necessary. Moreover, a fairly large portion of our needed software can be found in open source\textsuperscript{\cite{bep_0039,bep_0046,syncthing,tw_murder}} which will reduce needed resource to implement or test this concept.
\section{Cost Benefits Analysis}
\label{sec:org26f0ad0}
Here we presented the number of branches of only the \emph{state-owned banks, Bangladesh Election Commission} and \emph{public universities of Bangladesh} with a hypothetically least number of machines they run everyday. By machine we are referring to a computer with at least 2GB RAM, 500GB HDD and an average CPU. Servers or big machines are not included here.

\begin{center}
\begin{tabular}{lrrr}
\hline
Institutions & Branches & PCs & Total\\
\hline
Bangladesh Bank\textsuperscript{\cite{bb_branch}} & 10 & 50 & 500\\
Janata Bank Limited\textsuperscript{\cite{jb_branch}} & 914 & 5 & 4570\\
Agrani Bank Limited\textsuperscript{\cite{ab_branch}} & 956 & 5 & 4780\\
Rupali Bank Limited\textsuperscript{\cite{rb_branch}} & 672 & 5 & 3360\\
Sonali Bank Limited\textsuperscript{\cite{sb_branch}} & 1218 & 5 & 6090\\
Bangladesh Krishi Bank\textsuperscript{\cite{kb_branch}} & 1038 & 5 & 5190\\
Election Commission\textsuperscript{\cite{ec_dbs}} & 500 & 5 & 2500\\
Public Universities\textsuperscript{\cite{bd_pub_uni}} & 45 & 10 & 450\\
\hline
Total &  &  & 27440\\
\hline
\end{tabular}
\end{center}
This above number represent a tiny portion of machines run each day in government offices of \emph{Bangladesh}. Most of these are mainly used to browse web or do some word processing or spreadsheet things. So, if \emph{one percent} of these machines join our ID document verification private network, then we have,

\begin{center}
27440 \texttimes{} 1\% = 274.40 \(\approx\) 274 nodes
\end{center}
And if each node serve \texttt{100 requests/second} then together this network of nodes can handle,
\begin{center}
274 \texttimes{} 100 = 27400 requests/second
\end{center}
We admit that current \emph{Oracle Exadata Database Machine\textsuperscript{\cite{ec_oracle_exadata,ec_oracle_daily_star}}} based ID data management system of \emph{Bangladesh Election Commission} is more feature rich than our DIDB based system and we will not compare them. Rather we will analyze here the benefits of running them jointly.

\subsection{Economic Benefits}
\label{sec:orgdd88270}
As these nodes are consuming electricity anyway either we use hundred percent of them or not. So, we can run our ID verification system in background of these nodes with no extra cost. On the other hand by running it this way will make using ID document verification gateway like \emph{Porichoy\textsuperscript{\cite{porichoy_website}}} obsolete.

\subsection{Offline Verification}
\label{sec:org28a2e48}
This \emph{offline verification} is a quite interesting benefit of this system. In this network the nodes can go offline as they need. But they must need to get synced with the latest DIDB after coming online. As we said earlier, every node will response to verification requests by querying their local DIDB clone, hereby they will be able to do so in offline mode too when they get requests internally or from another node connected by LAN. Sometime we may need to verify identity document in a remote place or just after a natural disaster where communication is disrupted or at a place where getting internet is very difficult. In those cases we can use this system to verify the identity documents offline. But where ever we use this offline feature, we must need to ensure proper security and take counter measures that the DIDB inside the offline device does not get manipulated.

\subsection{Faster Verification}
\label{sec:org757661f}
As the data will be replicated in a number nodes, the verification will be much faster with almost no cost. And if any node is busy or unable to process the request, then the verification request can be forwarded to other private nodes. 

\subsection{High Sustainability}
\label{sec:org3b7c7c1}
In the gateway server model, if the gateway server or CIDB is busy or unable to serve requests, then nobody will get served. But in our DIDB based distributed system, these peers can act as server. Thus ,the probability of getting served all the time is very high. If any node is unable to handle request we can ask another node by choosing another IP address, which proves the sustainability of this distributed system.

\subsection{Isolation of CIDB}
\label{sec:org25016bc}
This system will make the use of CIDB for ordinary identity document verification obsolete. Therefore isolation of this central identity database or CIDB can be possible, which will enhance the security of the CIDB as well as the save the resource which would be needed for validating the ordinary identity document verification. 
\section{Conclusion}
\label{sec:org4d143ac}
Usually in \emph{Bangladesh}, people need to submit copy of an identity document where it is required. But verifying the content of that copied document while receiving it, often never done. It mostly happen because of either general practice or lake of a sustainable system. Thus there is still a probability left for fraud. By verifying the copy of ID documents at the time of receiving we can reduce this chance of fraud a lot. However, though we described this concept for ID document verification, it will work for verifying other identity documents like passport or driver's license or tax certificates too.

\bibliographystyle{unsrturl} 
\bibliography{khan_2020_bt_id_verify.bib}
\end{document}